\documentclass[prb, twocolumn, 10pt, final, floatfix, aps]{revtex4}
\usepackage{graphicx}
\usepackage{amsmath, amssymb, amsfonts, bm}
\usepackage{latexsym}
\usepackage[colorlinks]{hyperref}
\usepackage{epstopdf}

\begin{document}

\title[Dynamical Autler-Townes control of a phase qubit]{Dynamical Autler-Townes control of a phase qubit}

\author{Jian Li$^1$} \author{G.~S. Paraoanu$^1$} \author{Katarina Cicak$^2$} \author{Fabio Altomare$^{2\dag}$} \author{Jae I. Park$^2$}
\author{Raymond~W. Simmonds$^2$} \author{Mika~A. Sillanp\"a\"a$^1$} \author{Pertti J. Hakonen$^1$}


\affiliation{$^1$O.V. Lounasmaa Laboratory, Aalto University, PO Box 15100, FI-00076 AALTO, Finland \\
$^2$National Institute of Standards and Technology, 325 Broadway, Boulder, Colorado 80305, USA \\
$^\dag$Present address: D-Wave Systems Inc., 100-4401 Still Creek Drive, Burnaby, British Columbia V5C 6G9, Canada }


\maketitle


{\bf Routers, switches, and repeaters are essential components of modern information-processing systems.
Similar devices will be needed in future superconducting quantum computers. 
In this work we investigate experimentally the time evolution of Autler-Townes splitting in a superconducting phase qubit under  the application of a control tone resonantly coupled to the second transition. A three-level model that includes independently determined parameters for relaxation and dephasing gives excellent agreement with the experiment.  The results demonstrate that the qubit can be used as a ON/OFF switch with 100 ns operating time-scale for the reflection/transmission of photons coming from an applied probe microwave tone. The ON state is realized when the control tone is sufficiently strong to generate an Autler-Townes doublet, suppressing the absorption of the probe tone photons and resulting in a maximum of transmission.}



In the optical regime, a variety of effects that can be employed for switching have been demonstrated, for example photon blockade by single atoms \cite{photonblockade}, parametric instabilities in nonlinear optical wave-mixing \cite{parametricinstabilities}, heralded single-photon absorption by one trapped ion \cite{heralded}, single-atom dynamic control of light in microresonator photon turnstiles \cite{turnstiles}, and population inversion in single dye molecules \cite{transistor}. Optical switching using the phenomenon of electromagnetically induced transparency \cite{EIT} has been realized in gases of magneto-optically trapped $^{87}{\rm Rb}$ \cite{switch}; recently, it has become possible to use for this purpose only a single $^{87}{\rm Rb}$ atom in a high-finesse cavity \cite{rempe}. With improved dissipation and decoherence, it could be possible to use electromagnetic-induced transparency effects to build similar devices with quantum superconducting circuits \cite{astafiev}. Other applications employing multilevel systems - for example state preparation, emulation of large quantum-number spins \cite{neely}, single-shot fast quantum nondemolition readout techniques \cite{shelving}, implementation of control Z-$\pi$ gates in certain quantum computing architectures \cite{zpi}, single-qubit microwave amplifiers \cite{amp}, {\it etc.} - can be envisioned.

In this paper we demonstrate the dynamical operation of a phase qubit as a microwave photon-absorbing device that can be switched ON or OFF in approximately 100 ns by using an external control field. The functioning of our device is based on an effect closely related to electromagnetically induced transparency: under the application of a relatively intense continuous field (the control field) to the second transition of the qubit, the spectral line of the first transition splits into a doublet. This is called Autler-Townes effect \cite{AT}, and it has been recently observed in superconducting quantum systems as well \cite{wallraff,mika,bbn}. Here we operate our device in time-domain, which allows us to observe transient effects that occur until a steady state is reached. We observe how the doublet forms when the control field is suddenly applied, and how the spectral line of the first transition appears back from the merging of the Autler-Townes peaks when the coupling field is switched off.

Besides the fundamental importance for understanding the dynamics of the Autler-Townes effect, our experiment has potential applications in the field of microwave photonics, for example for fast quantum switches that can be integrated in superconducting quantum-processor architectures. This can be achived by realizing that the ``control'' microwave tone applied to the qubit determines the absorption rate for a ``probe'' microwave tone. Indeed, when the control field is OFF, the third level does not play any role and the device absorbs the incident probe field radiation resonant to the lowest transition, and its first excited level is populated. If such a two-level system is embedded into a transmission lines, it has been predicted theoretically in quantum optics \cite{shen} and demonstrated experimentally \cite{tsai} that the qubit will reflect back the incoming photons by resonant fluorescence, provided that the intensity of the probe field is not too high (this saturates the qubit). When the control field is ON, the absorption from the qubit is suppressed almost completely - the system becomes ``transparent'' to the incident probe radiation and the transmission is maximal. We calculate theoretically the expected reflection coefficient both in the transient and in the steady-state regimes, and we also analyze the performance of the switch at different strengths of the control and probe tones. Using our theoretical model, we are able to present a set of general quidelines for the design of superconducting switches based on the Autler-Townes effect - we show that high anharmonicity, low probe powers, and low dephasing rates lead to an effective suppressions of the reflection.

\section*{Results}

\begin{figure}
\includegraphics[width=1.0\linewidth]{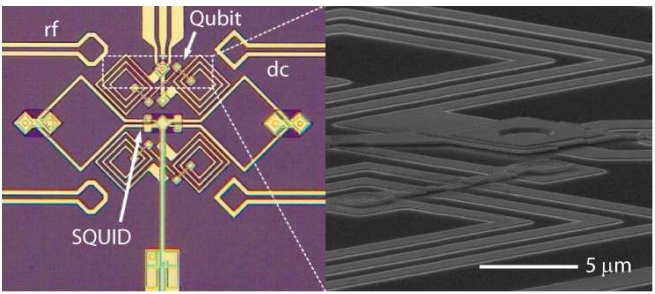}
\caption{{\bf Image of the Josephson phase qubit sample used in the experiment.} A ${\rm Al/AlO}_x{\rm /Al}$ junction with an area of $\sim 14~\mu{\rm m}^2$ was patterned on a sapphire substrate by standard optical lithography. ${\rm SiN}_x$ was used as an insulator between different metallic layers. }
\label{fig_sample}
\end{figure}

Our phase qubit \cite{singleshot} consists of an r.f. SQUID with loop inductance $L$, junction capacitance $C$, and Josephson energy $E_{J}$, which is read out by another d.c. SQUID fabricated on-chip in the proximity of the qubit and coupled inductively to it (see Fig.~\ref{fig_sample}). The qubit can be biased by an externally applied magnetic flux and can be coupled to other circuit elements such as transmission lines and resonators. The device forms a multilevel quantum system with the first three levels denoted by
$|0\rangle$, $|1\rangle$, and $|2\rangle$. These energy levels can be addressed using microwave fields closely resonant to the first $|0\rangle \rightarrow |1\rangle$ and the second $|1\rangle \rightarrow |2\rangle$ transitions (the
$|0\rangle \rightarrow |2\rangle$ transition is almost forbidden).

\begin{figure}
\includegraphics[width=1.0\linewidth]{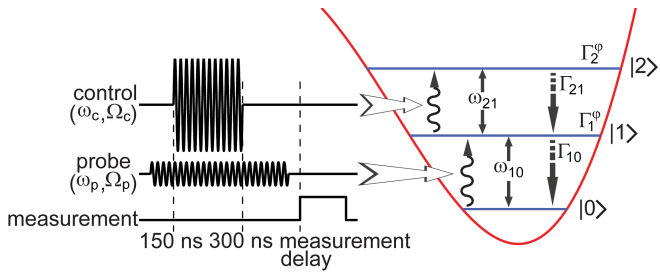}
\caption{{\bf Schematic of the pulse sequence for the dynamic Autler-Townes effect.} The qubit is probed by a continuous microwave field $(\omega_{p}, \Omega_{p})$, with the frequency $\omega_{p}$ being swept around the first frequency transition $\omega_{10}$. Each measurement sequence starts at $t=0$ ns with the system in the stationary state as given by the application of the probe field. At $t=150$ ns we apply the coupling pulse $(\omega_{c}, \Omega_{c})$, exactly resonant
with the second transition, $\omega_{c} = \omega_{21}$.
}
\label{fig_schematicdynamic}
\end{figure}

A detailed description of the functioning of the device has been given elsewhere \cite{jian}. The Hamiltonian of the system is
\begin{equation}
H = \frac{Q^2}{2C} + \frac{(\Phi - \Phi_{\rm ext})^2}{2L} - E_J\cos\left( 2\pi\frac{\Phi}{\Phi_0} \right) ,\label{eq_hamiltonian}
\end{equation}
where $Q$ is the charge on the capacitor formed by the junction, $\Phi$ is the flux variable, $\Phi_{0} = h/2e = 2.067 \times 10^{-15}$ Wb is the magnetic flux quantum, and  $\Phi_{\rm ext} = \Phi_{\rm dc} + \Phi_{\rm rf}(t)$ is the total magnetic-flux component of the externally applied d.c. and r.f. fields. In this system, the r.f. tones applied have angular frequencies $\omega_p$ (probe tone) and $\omega_c$ (control tone)
 close to the two resonant frequencies $\omega_{10}$ of $|0\rangle \rightarrow |1\rangle$ transition and $\omega_{21}$ of $|1\rangle \rightarrow |2\rangle$ transition, respectively (see Fig.~\ref{fig_schematicdynamic}). The first two transitions were determined in independent single-tone and two-tone spectroscopy measurements \cite{jian}, and they had the values $\omega_{10} = 2\pi \times 8.135$ GHz and $\omega_{21} = 2\pi \times 7.975$ GHz. The probe and control fields are supplied to the device through an on-chip coplanar waveguide transmission line. The Rabi frequencies of these fields are denoted by $\Omega_{p}$ and $\Omega_{c}$ respectively, and they are directly proportional to the corresponding probe and control field amplitudes.

\begin{figure}
\includegraphics[width=1.0\linewidth]{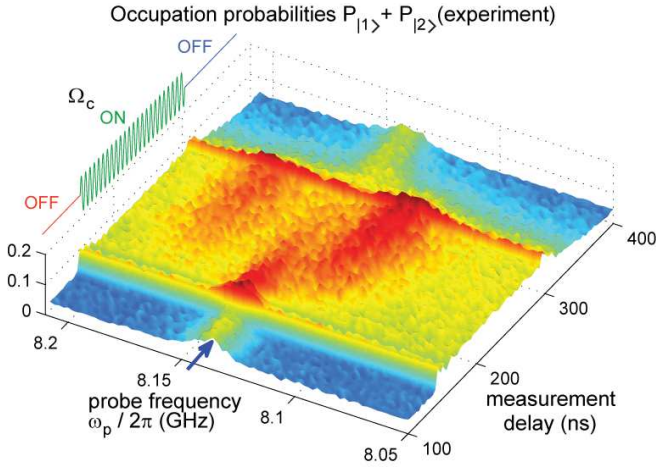}
\caption{{\bf On-set and  extinction of the Autler-Townes effect.} The figure shows the experimentally determined sum of excited-states level-occupancy probabilities, $P_{|1\rangle} + P_{|2\rangle}$, obtained by applying the pulse sequence from Fig.~\ref{fig_schematicdynamic}. The arrow indicates the frequency of the first transition, $\omega_{10}/2\pi = 8.135$ GHz.}
\label{fig_dynamic}
\end{figure}

{\bf Experimental data.}
Here we demonstrate that this three-level superconducting device can be operated dynamically, in time domain. 
In Fig.~\ref{fig_schematicdynamic} we present the pulse sequence applied in order to use the system as a switch based on the Autler-Townes effect.
In Fig.~\ref{fig_dynamic} we show the results of the experiment.
The ON state is characterized by the suppression of excitations to the state $|1\rangle$ normally caused by absorbtion of photons from the probe field. It is produced by a large value of the control field amplitude (effective Rabi frequency $\Omega_{c}$). The state OFF corresponds to a low enough value of the control field so that excitations to the state $|1\rangle$ are allowed. Although we do not measure the transmission (reflection) of photons directly, the data demonstrates unambiguously that the device can be used to modulate the transfer of photons by controlling the absorbtion rate of the qubit.
The measured state occupation numbers show clearly the dynamical process of formation of the Autler-Townes doublet, how it reaches the stationary state, and how the two peaks finally collapse into a single spectroscopic signal at the first transition frequency when the coupling field is switched off. In both the on-set and the switch-off stages of the effect, the timescale for the system to reach the steady state is of the order of 100 ns, in agreement with the decoherence times in our system (see the theoretical model below).

{\bf Numerical simulation.}
In order to simulate the experimental results including the effects of decoherence, we use the standard Liouville superoperators in the Markov approximation.
For our system, described by the Hamiltonian equation~(\ref{eq_hamiltonian}) and placed in a reservoir of electromagnetic modes at zero temperature (see \cite{jian} for a detailed derivation), the
Markovian master equation for the density matrix $\rho$ in the rotating frame reads
\begin{equation}
\dot{\rho} = -\frac{i}{\hbar}\left[\tilde{H}, \rho\right] + {\cal L}_{\rm rel}[\rho] + {\cal L}_{\rm dep}[\rho] . \label{eq_master_sp}
\end{equation}
The effective Hamiltonian $\tilde{H}$ is obtained by truncating equation~(\ref{eq_hamiltonian}) to the lowest three levels and performing a rotating wave approximation. It can be put in the form 
\begin{eqnarray}
\tilde{H} &=& \frac{\hbar}{2} \left[\Delta_{p}\sigma_{11} + (\Delta_{p} + \Delta_{c}) \sigma_{22}\right] + \nonumber \\
 && + \frac{\hbar}{2}  \left[(\Omega_{p}e^{i\phi} + 0.69\Omega_c e^{-i\delta t}) \sigma_{01}  \right. \nonumber \\
 && \ \ \ \ \ \ \left. + (\Omega_{c} + 1.45\Omega_p e^{i\delta t + i \phi}) \sigma_{12} \right]+ {\rm h.c.}. \label{eq_eff_hamiltonian}
\end{eqnarray}
Here we have taken an external r.f. magnetic  field of the form $\Phi_{\rm rf} (t) = \Phi_{p} \cos (\omega_{p} t + \phi) + \Phi_{c} \cos (\omega_{c} t)$. This produces an effective coupling to the
qubit $g (t) = g_{p} \cos (\omega_{p} t + \phi) + g_{c} \cos (\omega_{c} t)$, where $g_{c} = - (\Phi_{c}/L)\sqrt{1 /C\hbar \omega_{0}}$ and $g_{p} = - (\Phi_{p}/L)\sqrt{1/C\hbar \omega_{0}}$ (where $\omega_{0}$ is the Josephson plasma frequency)
resulting in Rabi frequencies $\Omega_{p} = 0.69 g_{p}$ and $\Omega_{c} = g_{c}$ (see Ref. \onlinecite{jian} for a detailed derivation).
Here we have used the operators $\sigma_{ij} \stackrel{\rm def}{=} |i\rangle \langle j|$ (which generalize
the Pauli matrices to a many-level system), and we have defined the detunings $\Delta_p = \omega_{10} - \omega_p$, $\Delta_c = \omega_{21} - \omega_c$, and $\delta = \omega_p - \omega_c$.
The origin of the qubit-field coupling term (the second line in the Hamiltonian equation~(\ref{eq_eff_hamiltonian}))can be understood easier if one treats the device as a harmonic oscillator interacting with fields; in this case, the corresponding terms in the Hamiltonian read $(\hbar /2)  \left[(1/\sqrt{2})(g_{p}e^{i\phi} + g_{c} e^{-i\delta t}) \sigma_{01}  + (g_{c} + g_{p} e^{i\delta t + i \phi}) \sigma_{12} \right]+ {\rm h.c.}$. The result of including anharmonic corrections is that $\sqrt{2}$ from the previous expression is replaced by $1.45$ and $1/\sqrt{2}$ by $0.69$.

The relaxation part is given by \cite{jian}
\begin{eqnarray}
{\cal L}_{\rm rel}[\rho] &=& \frac{\Gamma_{10}}{2}\left( 2\sigma_{01}\rho\sigma_{10} - \sigma_{11}\rho - \rho\sigma_{11} \right) \nonumber \\
 && + \frac{\Gamma_{21}}{2}\left( 2\sigma_{12}\rho\sigma_{21} - \sigma_{22}\rho - \rho\sigma_{22} \right) \nonumber \\
 && + \kappa \left( e^{-i\delta t} \sigma_{01}\rho\sigma_{21} + e^{i\delta t} \sigma_{12}\rho\sigma_{10} \right) ,
\label{eq_relax}
\end{eqnarray}
and the dephasing is
\begin{eqnarray}
{\cal L}_{\rm dep}[\rho] &=& \frac{\gamma^{\varphi}_{10}}{2}\left( 2\sigma_{11}\rho\sigma_{11} - \sigma_{11}\rho - \rho\sigma_{11} \right) \nonumber \\
&& + \frac{\gamma^{\varphi}_{20}}{2}\left( 2\sigma_{22}\rho\sigma_{22} - \sigma_{22}\rho - \rho\sigma_{22} \right).
\end{eqnarray}
Here $\rho$ is the density matrix in a doubly-rotating frame \cite{jian}, the interlevel relaxation rates between $|1\rangle \rightarrow |0\rangle$ and $|2\rangle \rightarrow |1\rangle$ are denoted as $\Gamma_{10}$ and $\Gamma_{21}$, respectively, $\gamma_{10}^{\varphi}$, $\gamma_{20}^{\varphi}$ are intrinsic (pure) dephasing rates of the states $|1\rangle$ and $|2\rangle$, and $\kappa = \sqrt{\Gamma_{10}\Gamma_{21}}$. The dissipation parameters were either measured directly in independent experiments ({\it e.g.} by exciting the qubit and measuring the decay time) or extracted from spectroscopy data \cite{mika, jian}, and for this sample they were
$\Gamma_{21} = 2\pi\times 11$ MHz, $\Gamma_{10} = 2\pi\times 7$ MHz, $\gamma^{\varphi}_{10} = 2\pi\times 7$ MHz,
 $\gamma^{\varphi}_{20} = 2\pi\times 16$ MHz.

\begin{figure}
\includegraphics[width=1.0\linewidth]{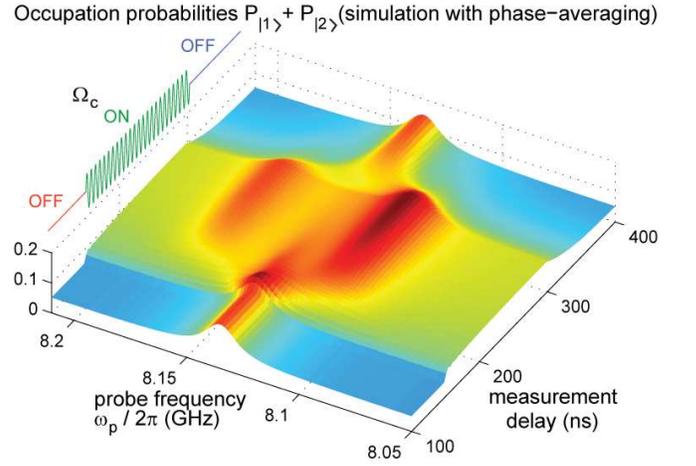}
\caption{{\bf Simulated occupation probabilities for  the sum of the occupation probabilities for the states
$|1\rangle$ and $|2\rangle$ with averaging over the phases $\phi$.}
This quantity is measured directly in the experiment (see Fig.~\ref{fig_dynamic}).}
\label{probabilities}
\end{figure}

In Fig.~\ref{probabilities} we present the result of the numerical simulation using equation~(\ref{eq_master_sp}).
For this simulation, the amplitude of the probe field corresponds to Rabi frequencies $\Omega_p = 2\pi \times 3.45$ MHz; for the control field, the corresponding values were $\Omega_c=2\pi \times 60$ MHz in the ON state, and a much smaller value $\Omega_c=2\pi \times 5$ MHz in the OFF state. The calibration of the probe and control field amplitudes was done in independent measurements of the Rabi frequency between the levels $|0\rangle$ and $|1\rangle$, and between the levels $|1\rangle$ and $|2\rangle$ respectively. In the absence of the control field, the level $|2\rangle$ is empty while the level $|1\rangle$ is populated when the probe field is resonant with the $|0\rangle \rightarrow |1\rangle$ transition. After the control field has been switched on, the original high occupation peak centered at $\omega_{10}$ splits into the Autler-Townes doublet. These observations are in good agreement with the experiment (Fig.~\ref{fig_dynamic}).  The simulation also captures two additional features of the experimental data: the asymmetry in the  Autler-Townes peaks and the appearance of a plateau
with nonzero occupation probabilities even when $\omega_p$ is far off-resonant from $\omega_{10}$ in the ON state. Both of these effects result from the existence of cross-coupling terms in the Hamiltonian equation~(\ref{eq_eff_hamiltonian}), see the Discussion below.

In the experiment, the phase difference $\phi$ between the probe and the control field cannot be maintained at the same value for each realization of the sequence presented in Fig.~\ref{fig_schematicdynamic}. To model this, we calculate the probabilities $P_{|1\rangle} + P_{|2\rangle}$ for several values of $\phi$ randomly distributed in the interval $[0,2\pi ]$ and we average over the results. More considerations about the effect of the cross-coupling terms are delegated to the Discussion.

\begin{figure*}
\includegraphics[width=1.0\linewidth]{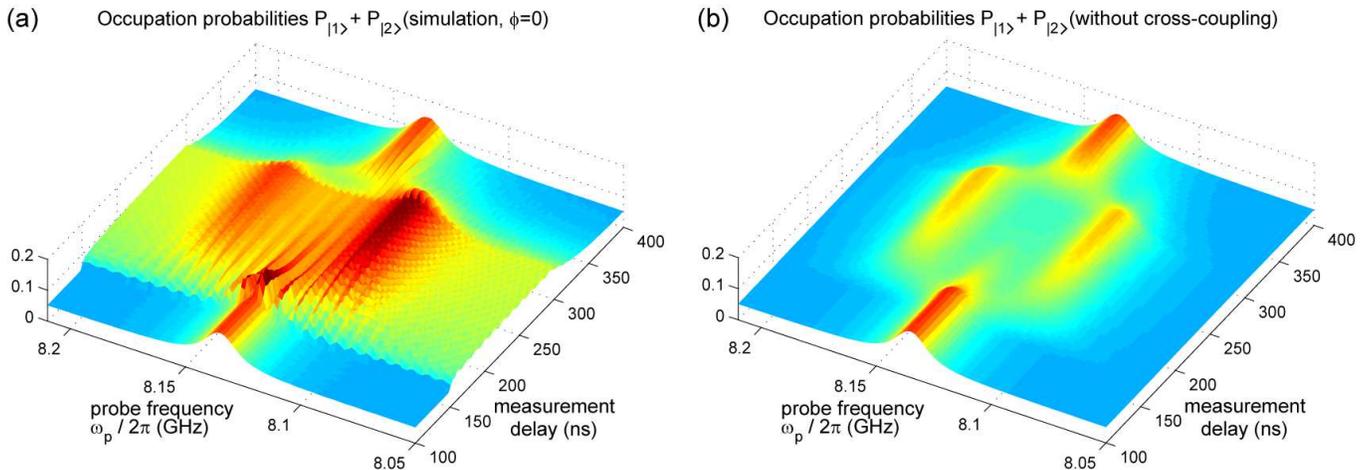}
\caption{{\bf Simulated 
$P_{|1\rangle} + P_{|2\rangle}$ with $\phi =0$, and (a) with cross-coupling, (b) without cross-coupling. }
}\label{fig_zerophase}
\end{figure*}

We note that the power levels used in this experiment correspond to operation close to the level of single microwave photons. To estimate the average number of  photons involved, consider a generic process (without dissipation) of resonant Rabi oscillations with Rabi frequency $\Omega_{\rm R}$. The probability for a two-level system to be excited in a time $\tau$ is $\sin^{2}(\Omega_{\rm R}\tau /2)$. Therefore, after a time $\tau_{R} = \pi /\Omega_{\rm R}$ the system is with certainty in the first excited state, meaning that it has absorbed exactly one photon. The corresponding power is $\hbar \omega /
\tau_{p}$. In our case, the power levels correspond to a Rabi frequency $\Omega_{p} = 2\pi \times 3.5$ MHz, resulting in $\tau_{p} = 143$ ns (and a power of $\hbar\omega_{10}/\tau_{p} = 38$ aW = -134 dBm), and for the
control field in the ON state we have $\Omega_{c} = 2\pi \times 60$ MHz, yielding $\tau_{c} = 8.3$ ns (and a power of
$\hbar\omega_{21}/\tau_{c} = 0.6$ fW = -122 dBm).
Thus, during the time between $t=150$ ns and $t=300$ ns there were on average about 1 photon present in the probe beam and
about 18 photons in the control beam. 
Moreover, as shown in the next section, increasing the input power level saturates the qubit and switching does not occur anymore. These low power levels are typical for measurements of superconducting qubits, therefore the devices based on the mechanism described in this paper can be seamlessly integrated in any circuit QED architecture.


\section*{Discussion}

{\bf The effect of the cross coupling terms.}
In the effective Hamiltonian equation~(\ref{eq_eff_hamiltonian}), there are oscillating terms with factors $\exp(\pm i\delta t)$, which result from cross couplings of the coupling field into the $|0\rangle \rightarrow |1\rangle$ transition and of the probe field into the $|1\rangle \rightarrow |2\rangle$ transition (see also the detailed derivation in Sec. II and Sec. III of Ref. \onlinecite{jian}). In the real experiment, it was not possible to ensure that the phase between the coupling and the probe fields remained the same for every run of the measurement sequence. To obtain Fig.~\ref{probabilities}, we have averaged over $\phi$, which is what occurred effectively in the experiment as well. It is however instructive to see what would happen for a constant phase (or a single-shot measurement), and for the case in which there is no cross-coupling at all.

In Fig.~\ref{fig_zerophase}(a) we show the result of simulations with the same master equation as equation~(\ref{eq_master_sp}) but with the relative phase between the probe and the coupling fields fixed at $\phi = 0$.
One can see the appearance of fringe-like structures in the spectrum. This figure can be regarded as a snapshot corresponding to a single run of the pulse sequence. The plateau seen in the previous spectra in the region where the coupling field is on and the probe is far off-resonant still exists; also the asymmetry between the Autler-Townes peaks is preserved. At different $\phi$'s, these fringes appear in different places; thus when averaging over $\phi$ they will be washed away.

To get even more understanding of the effects of the coupling terms, let us analyze what happens if  there is no cross-coupling at all. This would be the case for a strongly anharmonic system. In this case we can approximate the effective Hamiltonian as
\begin{equation}
\tilde{H}' = \frac{\hbar}{2} \left[\Delta_{p}\sigma_{11} + (\Delta_{p} + \Delta_{c}) \sigma_{22}\right] +
  \frac{\hbar}{2}  \left[\Omega_{p}  \sigma_{01}  + \Omega_{c} \sigma_{12} \right]+ {\rm h.c.}
\end{equation}
This Hamiltonian is widely used in  quantum optics for studying electromagnetically induced transparency and coherent population trapping \cite{EIT}.

Fig.~\ref{fig_zerophase}(b) shows a time-domain simulation similar to that in Fig.~\ref{fig_zerophase}(a) but with the Hamiltonian $\tilde{H}'$. The time-oscillating terms in ${\cal L}_{\rm rel}[\rho]$ are similarly neglected.
Several features can be noticed immediately. The Autler-Townes peaks are this time symmetric, the occupation probabilities at $\omega_p$ far away from Autler-Townes splittings are smaller, and we don't have a plateau anymore in the ON state (from 150 ns to 300 ns). The latter two features can be  explained intuitively in a simple manner: the coupling field tends to introduce an additional population of level $|1\rangle$ when it couples into the $|0\rangle \rightarrow |1\rangle$ transition. This effect does not depend on the frequency of the probe field, which is why it is appears as a plateau.


{\bf Operation as a quantum switch.}

The capability of distributing entangled microwave photons in superconducting  quantum networks is an essential ingredient for building future quantum processors \cite{njp}. An immediate application of the effect presented above is to realize a quantum switch for controlled routing of microwave photons. These devices can be realized by embedding the qubit into an open transmission line and using a vector network analyzer to measure the scattering coefficients of the probe field, yielding the corresponding reflection and transmission coefficients. In the stationary regime, these measurements have been done for flux qubits \cite{astafiev} and transmons \cite{chalmers}. Here we discuss the general dynamic characteristics of these devices, with exemplification by using the parameters corresponding to our qubit.


For qubits embedded into one-dimensional transmission lines, the reflection coefficient $r$ of the probe tone is proportional to the off-diagonal element $\rho_{10} = {\rm Tr} [\rho \sigma_{10}]$\cite{astafiev}:
\begin{equation}
r = r_{0} \frac{i \Gamma_{10}}{\Omega_p} \rho_{10}, \label{refl}
\end{equation}
where $r_0$ is the maximum reflection amplitude \cite{tsai}. Note that two conventions are used in the literature, one that considers the reflection as positive and the other in which it is taken negative.
Our equations do not depend on the convention used, but we assume that the same convention is applied both to $r$ and to $r_0$. In the case of our experiment, the switching occurs between two steady states, $\rho^{\rm (st)}_{\rm ON}$ and $\rho^{\rm (st)}_{\rm OFF}$, which are reached after $\approx$100 ns of transients. As a figure of merit for switches, the OFF/ON power ratio between the reflected powers
in the OFF state and that in the ON state $R_{\rm OFF}/R_{\rm ON}$
must be as large as possible. In the simulations below we will use parameters corresponding to our experiment; however, we stress that the theoretical results are generic and can be applied to any type of superconducting quantum multi-level system.

\begin{figure*}
\includegraphics[width=0.9\linewidth]{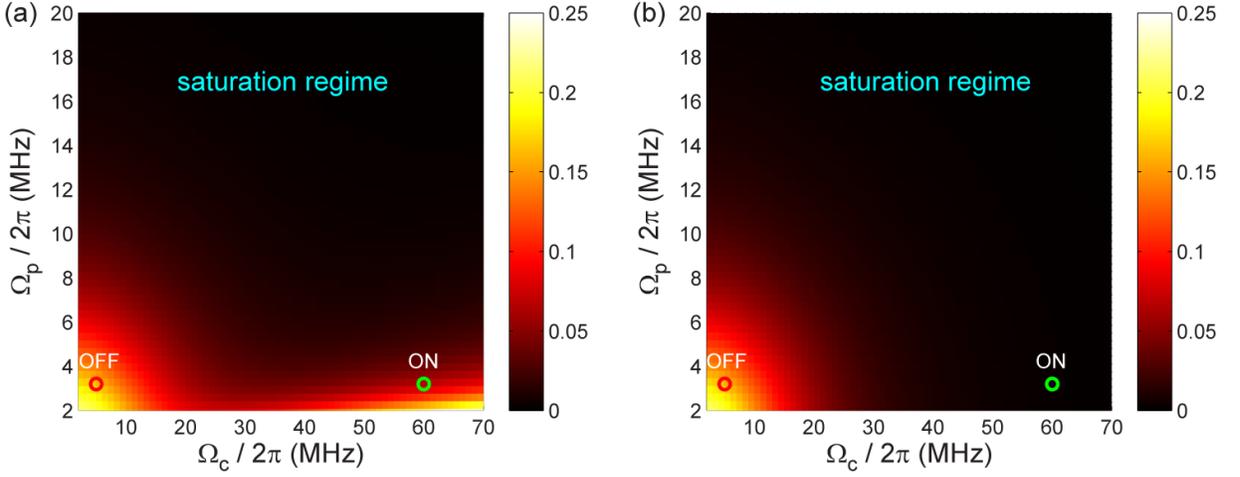}
\caption{{\bf On-resonant reflected power in the steady state (a) with cross-coupling, and (b) without cross-coupling.}
The two circles indicate the steady states $\rho_{\rm OFF}^{\rm (st)}$ (red) and $\rho_{\rm ON}^{\rm (st)}$ (green) corresponding to our experiment. The decoherence rates are $\gamma_{10} = 2\pi\times 14$ MHz and $\gamma_{20} = 2\pi\times 27$ MHz. }\label{steady}
\end{figure*}

{\it Cross-coupling terms and anharmonicity}
We first calculate numerically the off-diagonal density matrix element $\rho_{10}^{(st)}$ and the reflection coefficient
obtained from equation \ref{refl} in the steady state. For these calculations we take both the probe and control fields resonant with the respective transitions ($\Delta_p = 0$ and $\Delta_c = 0$). This corresponds to the experimental data along the arrow from Fig.~\ref{fig_dynamic}. In Fig.~\ref{steady}(a) and (b) we present the normalized reflected power $R=|r/r_{0}|^2$ simulated with cross-coupling (using the Hamiltonian $\tilde H$), and respectively without cross-coupling (using the Hamiltonian $\tilde H'$). The effect of the  cross-coupling terms become apparent especially
at high values of $\Omega_{c}$; as this field couples into the $|0\rangle- |1\rangle$ transition, the reflected power is not turned down completely. We have checked numerically that this is due to the real part of $\rho_{10}^{(st)}$ acquiring a non-zero value, while the imaginary parts of $\rho_{10}^{(st)}$ with and without cross-coupling do not differ much. For example, without cross-coupling, ${\rm Re}[\rho_{10}^{(st)}] = 0$, ${\rm Im}[\rho_{10}^{(st)}] \approx -0.19$ in the OFF state, and ${\rm Re}[\rho_{10}^{(st)}] = 0$, ${\rm Im}[\rho_{10}^{(st)}] \approx -0.02$ in the ON state, which give a  power ratio $R_{\rm OFF}/R_{\rm ON}$ of 90; whereas with cross-coupling,  ${\rm Re}[\rho_{10}^{(st)}] \approx -0.01$, ${\rm Im}[\rho_{10}^{(st)}] \approx -0.19$ in the OFF state, and ${\rm Re}[\rho_{10}^{(st)}] \approx -0.1$, ${\rm Im}[\rho_{10}^{(st)}] \approx -0.04$ in the ON state, which give a ratio $R_{\rm OFF}/R_{\rm ON}$ of only 3.

This means that, in the state ON, the normalized power reflection $R=|r/r_0|^2$ with cross-coupling is always larger than that without cross-coupling, and the additional power reflection due to the cross-coupling reduces the on/off ratio, $R_{\rm OFF} / R_{\rm ON}$, of the switch. We conclude that, in order to have an efficient switch, it is advantageous to increase the anharmonicity of the qubit,  which would reduce the cross-couplings (for example in Eq. (\ref{eq_eff_hamiltonian}), if $\delta$ is large, the cross-coupling terms will oscillate fast and average to zero). Some types of qubits, for example the flux qubit, do have a high degree of anharmonicity by design; for phase qubits there exist as well proposals for increasing the anharmonicity \cite{Zorin}.

{\it Saturation}  If the probe power is too large, from 
Fig.~\ref{steady} we can see that, in the OFF state, the power reflection $R$ is significantly reduced. This is because in the OFF state the system is essentially a two-level system exposed only to the probe field radiation. In this case, if  $\Omega_p$ is much larger than $\gamma_{10}$, the stationary state of the system in the $\{|0\rangle ,|1\rangle$ subspace is maximally mixed, and the reflection coefficient drops to zero. This shows that increasing the power of the probe field would not improve the characteristics of a switch.

\begin{figure*}
\includegraphics[width=0.9\linewidth]{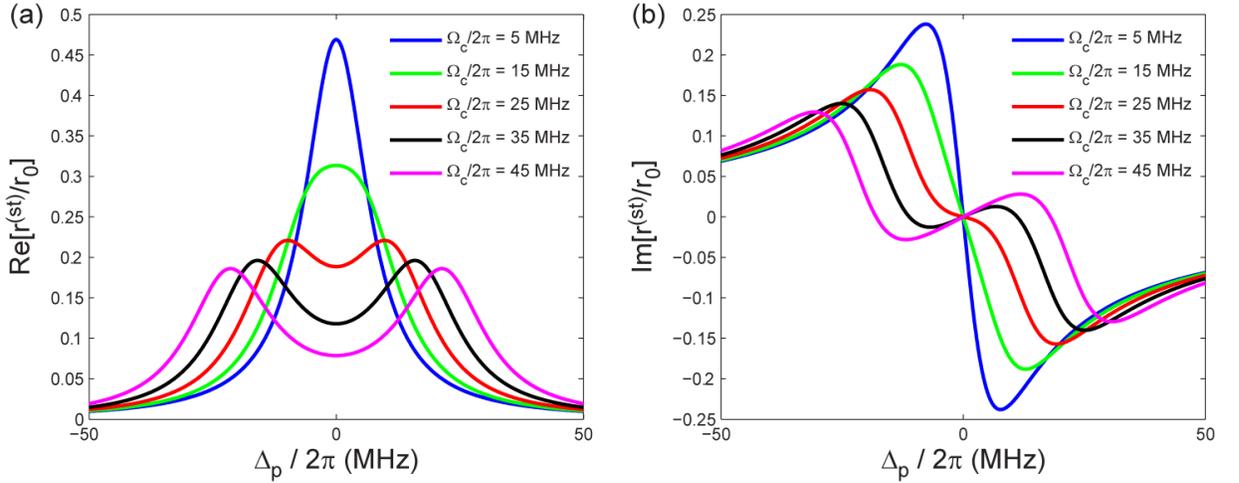}
\caption{{\bf (a) Real part and (b) imaginary part of the reflection coefficient as a function of detuning $\Delta_p$, at different control field amplitudes $\Omega_c$.} }\label{fig_rho_10}
\end{figure*}


{\it Dephasing}
In the absence of cross-couplings and for small enough values of  the probe field, $\left(\Omega_p/\gamma_{10}\right)^2\ll 1$ and $\left(\Omega_p / \Omega_c \right)^2\ll 1$, analytical results for 
for the off-diagonal element $\rho_{10}^{({\rm st})}$ 
in the steady-state are available \cite{jian},
\begin{equation}
\rho_{10}^{({\rm st})} = \frac{\Omega_p (2\Delta_p - i\gamma_{20})}{-4\Delta_p^2 + \Omega_c^2 + \gamma_{10}\gamma_{20} + 2i\Delta_p(\gamma_{10} + \gamma_{20})} , \label{eq_rho_10}
\end{equation}
where $\gamma_{10} = \Gamma_{10} + \gamma_{10}^\varphi = 2\pi \times 14$ MHz and $\gamma_{20} = \Gamma_{21} + \gamma_{20}^\varphi = 2\pi \times 27$ MHz.
In Fig.~\ref{fig_rho_10}, the real and imaginary parts of the reflection coefficient obtained with $\rho_{10}^{({\rm st})}$ of equation (\ref{eq_rho_10}) are plotted for different values of $\Omega_c$.


\begin{figure}
\includegraphics[width=1.0\linewidth]{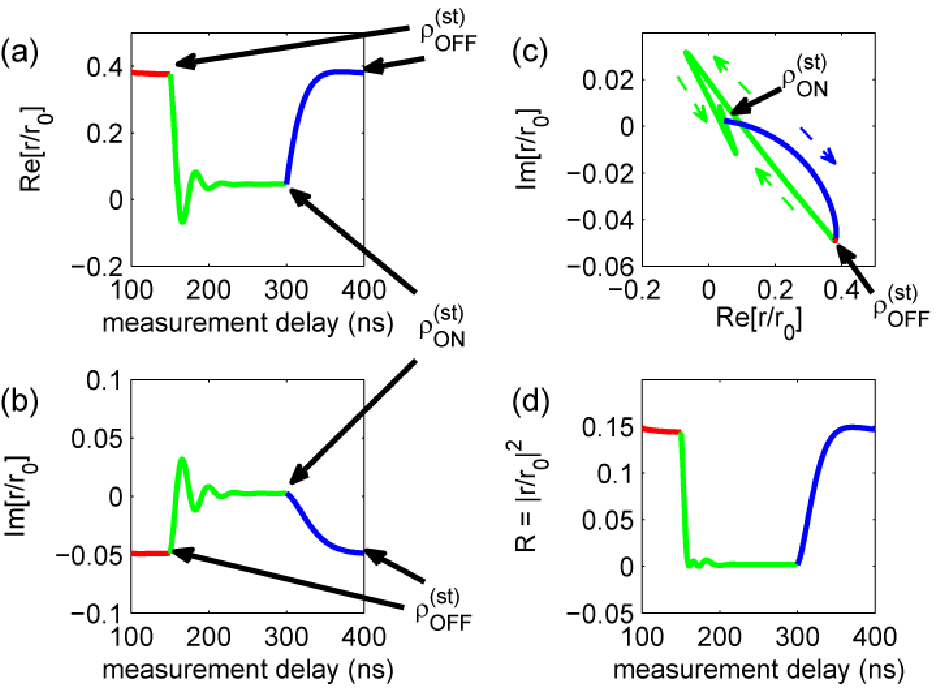}
\caption{{\bf (a) Real part, (b) imaginary part, and (c) Smith chart of the normalized reflection coefficient in time domain. (d) The corresponding (normalized) power reflection.} The control field is OFF before the 150 ns delay time (red), then switched ON between 150 and 300 ns (green), and finally switched OFF again (blue). The dashed arrows in (c) indicate the direction of time-evolution. The simulations correspond to the experimental data along the arrow from Fig.~\ref{fig_dynamic}.}\label{fig_reflection}
\end{figure}

From equation~(\ref{eq_rho_10}) we can also obtain an explicit expression for the reflection coefficient $r$ in the steady
state at probe-field resonance, 
\begin{equation}
r^{\rm (st)} = r_0 \frac{\Gamma_{10}\gamma_{20}}{\Omega_c^2 + \gamma_{10}\gamma_{20}} .  
\label{eq_r2}
\end{equation}
Using equation~(\ref{eq_r2}) we find that in the OFF steady state the reflection coefficient is $r^{\rm (st)}_{\rm OFF}/r_0 \approx 0.47$. Ideally, the reflection coefficient could approach the unit, but for the parameters corresponding to our qubit this does not happen due to the relatively large dephasing rate, $\gamma_{10} = 2 \Gamma_{10}$.  Now, when the coupling field is ON and the system reached the stationary state $\rho_{\rm ON}^{\rm (st)}$, equation~(\ref{eq_r2}) predicts that
there is almost no reflection ($r^{\rm (st)}_{\rm ON}/r_0 < 0.05$) of the probe field (green circle in Fig.~\ref{steady}).
These results are in reasonably good agreement with the numerical simulations in time domain
(see also Fig.~\ref{fig_reflection}), even though the approximation $\left(\Omega_p /\Omega_c \right)^2\ll 1$ is
not valid {\it stricto sensu}. For example, equation~(\ref{eq_r2}) predicts that the ON/OFF ratio $R_{\rm OFF}/R_{\rm ON}$
is of  the order of 100, in agreement with the numerical result obtained before. To obtain a higher reflection in the OFF state, equation~(\ref{eq_r2}) shows that one needs to have a lower dephasing rate and to reduce the value of the coupling field in the OFF state. If the coupling field in the OFF state is well below the value of the decoherence rates, we have from equation (\ref{eq_r2}) that $R_{\rm OFF}/R_{\rm ON} \approx \left(1 + \Omega_{c {\rm (ON)}}^2/\gamma_{10}\gamma_{20}\right)^2 $. As noticed before, increasing $\Omega_{c {\rm (ON)}}$ is ultimately limited by anharmonicity. Experimentally, lowering the OFF field to zero is limited by the leakage of the mixer used to pulse the coupling field, while the reduction of dephasing is a general problem in the field of superconducting qubits.


{\it Time-domain transients}
In Fig.~\ref{fig_reflection} we present the results of time-domain numerical simulations of the reflection coefficients for a switching event corresponding to the one occurring in the experiment presented in Fig.~\ref{fig_dynamic}. We see that, as in the case of the population number, the transients tend to stabilize after times of the order of 100 ns.

To conclude, our results demonstrate that a superconducting three-level system with large enough anharmonicity and low dephasing can be used for example as the node of a quantum network, redirecting single photons along chosen paths (reflecting them in the OFF state and allowing them to pass in the ON state).

\section*{Methods}


To probe the qubit we use a weak continuous probe tone. The Hamiltonian equation~(\ref{eq_hamiltonian}) is dynamically manipulated by changing the amplitude of the control field, resonantly coupling the two higher levels $|1\rangle$ and $|2\rangle$. This field is turned on at 150 ns and turned off at 300 ns. For measuring the occupation probabilities, we use the fact that the three levels reside in a metastable well \cite{singleshot}. The potential barrier that creates this well is then lowered by a measurement pulse applied after a measurement delay time (see Fig.~\ref{fig_schematicdynamic}), such that the states above the ground state are allowed to tunnel out of the well with a certain tunneling probability. The resulting variation in the magnetic flux piercing the qubit loop is detected by a nearby dc SQUID, which is switched in the running state \cite{me}. Therefore from the switching statistics we can calculate the total state occupation probability of the two excited levels $P_{|1\rangle }+ P_{|2\rangle}$ \cite{singleshot} that have tunneled out. The numerical simulations are done by solving the master equation with the fourth-order Runge-Kutta method \cite{Press}.




\section*{Acknowledgements}
We acknowledge financial support from the Academy of Finland (nos. 129896, 118122, 130058, 135135, and 141559), from the National Graduate School of Material Physics, from NIST, and from the European Research Council  (StG).


\section*{Contributions}
The samples were designed and fabricated by the authors affiliated with  NIST (K.C., F.A.,J.I.P., and R.W.S.). The measurements, the development of the theoretical model, the data analysis, and the writing of the manuscript was done by the authors affiliated with Aalto university (J.L., G.S.P., M.A.S., and P.J.H.). All authors discussed the results.



\begin{thebibliography}{99}




\bibitem{photonblockade} K. M. Birnbaum, A. Boca, R. Miller, A. D. Boozer, T. E. Northup and H. J. Kimble, Nature {\bf 436}, 87-90 (2005).


\bibitem{parametricinstabilities} A. M. C. Dawes, L. Illing, S. M. Clark, and D. J. Gauthier, Science {\bf 308}, 672-674 (2005).



\bibitem{heralded} N. Piro, F. Rohde, C. Schuck, M. Almendros, J. Huwer, J. Ghosh, A. Haase, M. Hennrich, F. Dubin, J. Eschner, Nature Physics {\bf 7}, 17 (2011).


\bibitem{turnstiles}
B. Dayan, A. S. Parkins, T. Aoki, E. P. Ostby, K. J. Vahala, and H. J. Kimble,
Science {\bf 319}, 1062 (2008).



\bibitem{transistor} J. Hwang, M. Pototschnig, R. Lettow, G. Zumofen, A. Renn, S. G\"otzinger, and V. Sandoghdar, Nature {\bf 460}, 76 (2009).


\bibitem{EIT} For a review, see M. Fleischhauer, A. Imamoglu, and J. P. Marangos, Rev. Mod. Phys. {\bf 77}, 633 (2005).



\bibitem{switch}
D. A. Braje, V. Balic, G. Y. Yin, and S. E. Harris, Phys. Rev. A {\bf 68}, 041801 (2003).



\bibitem{rempe} M. M\"ucke, E. Figueroa, J. Bochmann, C. Hahn, K. Murr, S. Ritter, C. J. Villas-Boas, G. Rempe, Nature {\bf 465}, 755 (2010); see also D. Castelvecchi,
Scientific American {\bf 303}, 26 (2010).



\bibitem{astafiev} A. A. Abdumalikov Jr., O. Astafiev, A. M. Zagoskin, Yu. A. Pashkin, Y. Nakamura, and J. S. Tsai, Phys. Rev. Lett. {\bf 104}, 193601 (2010).

\bibitem{neely} M. Neeley, M. Ansmann, R. C. Bialczak, M. Hofheinz, E. Lucero, A. D. O'Connell, D. Sank, H. Wang, J. Wenner, A. N. Cleland, M. R. Geller, and J. M. Martinis, Science  {\bf 325}, 722 (2009).

\bibitem{shelving} B. G. U. Englert, G. Mangano, M. Mariantoni, R. Gross, J. Siewert, and E. Solano, Phys. Rev. B {\bf 81}, 134514 (2010).

\bibitem{zpi} M. Mariantoni, H. Wang, T. Yamamoto, M. Neeley, R. C. Bialczak, Y. Chen, M. Lenander, E. Lucero, A. D. O'Connell, D. Sank, M. Weides, J. Wenner, Y. Yin, J. Zhao, A. N. Korotkov, A. N. Cleland, and J. M. Martinis, Science {\bf 334}, 61 (2011).


\bibitem{amp} O. Astafiev, A. A. Abdumalikov Jr., A. M. Zagoskin, Yu. A. Pashkin, Y. Nakamura, and J. S. Tsai, Phys. Rev. Lett. {\bf 104}, 183603 (2010).

\bibitem{AT} S. H. Autler and C. H. Townes, Phys. Rev. {\bf 100}, 703
(1955).

\bibitem{wallraff}
M. Baur, S. Filipp, R. Bianchetti, J. M. Fink, M. G\"oppl, L. Steffen, P. J. Leek, A. Blais, and A. Wallraff, Phys. Rev. Lett. {\bf 102}, 243602 (2009).

\bibitem{mika}
M. A. Sillanp\"a\"a, J. Li, K. Cicak, F. Altomare, J. I. Park, R. W. Simmonds, G. S. Paraoanu, and P. J. Hakonen, Phys. Rev. Lett. {\bf 103}, 193601 (2009).

\bibitem{bbn}
W. R. Kelly, Z. Dutton, J.  Schlafer, B. Mookerji, T. A. Ohki,
J. S. Kline and D. P. Pappas, Phys. Rev. Lett. {\bf 104}, 163601 (2010).

\bibitem{shen} J.T. Shen and S. Fan, Opt. Lett. {\bf 30}, 2001 (2005).

\bibitem{tsai} O. Astafiev, A. M. Zagoskin, A. A. Abdumalikov Jr., Yu. A. Pashkin, T. Yamamoto, K. Inomata, Y. Nakamura, and J. S. Tsai, Science {\bf 327}, 840-843 (2010).

\bibitem{singleshot} J. M. Martinis, S. Nam, J. Aumentado, and C. Urbina,
Phys. Rev. Lett. {\bf 89}, 117901 (2002); K. B. Cooper, M. Steffen, R. McDermott, R. W. Simmonds, S. Oh,  D. A. Hite,
D. P. Pappas, and J. M. Martinis, Phys. Rev. Lett. {\bf 93}, 180401
(2004).


\bibitem{jian}
J. Li, G. S. Paraoanu, K. Cicak, F. Altomare, J. I. Park, R. W. Simmonds, M. A. Sillanp\"a\"a, and P. J. Hakonen, Phys. Rev. B 84, 104527 (2011).


\bibitem{njp} J. Li and G. S. Paraoanu, New J. Phys. {\bf 11}, 113020 (2009).

\bibitem{chalmers} I.-C. Hoi, C. M. Wilson, G. Johansson, T. Palomaki, B. Peropadre, and P. Delsing, Phys. Rev. Lett. {\bf 107}, 073601 (2011).


\bibitem{nielsen} M. A. Nielsen and I. L. Chuang, Quantum Computation and Quantum Information (Cambridge University Press, Cambridge, 2000).

\bibitem{me} See {\it e.g.} G. S. Paraoanu, Phys. Rev. B {\bf 72}, 134528
(2005).

\bibitem{Zorin} 
A. B. Zorin and F. Chiarello, Phys. Rev. B {\bf 80}, 214535 (2009).


\bibitem{Press} W. H. Press, S. A. Teukolsky, W. T. Vetterling, and B. P. Flannery, Numerical Recipes, 3rd ed. (Cambridge University Press, Cambridge, 2007).






\end{thebibliography}
\end{document}